# Direct observation of number squeezing in an optical lattice

A. Itah, H. Veksler, O. Lahav, A. Blumkin, C. Moreno, C. Gordon and J. Steinhauer

*Department of Physics, Technion – Israel Institute of Technology, Technion City, Haifa 32000, Israel*

We present an in-situ study of an optical lattice with tunneling and single lattice site resolution. This system provides an important step for realizing a quantum computer. The real-space images show the fluctuations of the atom number in each site. The sub-Poissonian distribution results from the approach to the Mott insulator state, combined with the dynamics of density-dependent losses, which result from the high densities of optical lattice experiments. These losses are clear from the shape of the lattice profile. Furthermore, we find that the lattice is not in the ground state despite the momentum distribution which shows the reciprocal lattice. These effects may well be relevant for other optical lattice experiments, past and future. The lattice beams are derived from a microlens array, resulting in lattice beams which are perfectly stable relative to one another.

When loaded with atoms from a Bose-Einstein condensate, the optical lattice with tunneling between sites has proven to be a rich system. It allowed for the study of matter wave interference[1] and the superfluid to Mott-insulator transition[2-5] including squeezed number states[6,7], shells of definite atom number[8,9], and the disappearance of superfluid flow[10]. An optical lattice serves as an excellent model for condensed matter systems, and has been studied in great depth[11,12]. Furthermore, an optical lattice is a candidate for quantum computing[13], where single atoms in each site would be addressed individually. However, most optical lattice experiments to date have observed the lattice in momentum space[4-6], or have observed the overall profile of the lattice[8,9,14]. We present an in-situ study of an optical lattice with single lattice site resolution. This system provides an important step for realizing a quantum computer. The real-space images show the fluctuations of the atom number in each site. The sub-Poissonian distribution results from the approach to the Mott insulator state, combined with the dynamics of density-dependent losses, which result from the high densities of optical lattice experiments[4,8,15]. These losses are clear from the shape of the lattice profile. Such losses can potentially be used to prepare interesting quantum phases[16]. Furthermore, we find that the lattice is not in the ground state despite the momentum distribution which shows the reciprocal lattice. These effects may well be relevant for other optical lattice experiments, past and future. The lattice beams are derived from a microlens array, resulting in lattice beams which are perfectly stable relative to one another.

In a ground-state optical lattice with a negligible on-site repulsive interaction energy $U$ relative to the hopping energy $J$, the wavefunction of each atom spreads across the entire lattice, giving the same phase to each lattice site. The uncertainty in the phase for each lattice site $\sigma_\phi$ is much less than unity. Due to the small interaction energy, the position of each atom is independent of the other atoms, resulting in a Poissonian distribution of the number of atoms per site $N_i$, with a width $\sigma_N = \sqrt{N}$, where $N$ is the average number of atom per site. For larger values of $U/J$, $\sigma_N$ and $\sigma_\phi$ vary slowly as $(U/J)^{-1/4}$ and $(U/J)^{1/4}$, respectively[17]. For even larger values of $U/J$, the Mott transition is reached, where $\sigma_N = \sigma_\phi \approx \sqrt{2/\pi}$. We directly observe $\sigma_N$ by counting the number of atoms in each lattice site.

However, optical lattices often exhibit significant losses, at higher rates than expected[8,15,18,19]. Such losses can change $\sigma_N$ (see Appendix). If the loss rate $\tau_m^{-1}$ is large compared to the tunneling rate $J/\hbar$, then $\sigma_N$ will be dominated by the losses. In this case, the uncertainty in $N$ obtains the value $\sigma_N \approx \sqrt{5N/[4(m-1)+10]}$, for losses due to collisions between $m$ atoms. For 3-body losses for example, $\sigma_N \approx 0.53\sqrt{N}$. In our experiment, this width places a lower limit on $\sigma_N$ as the transition to the Mott insulator is approached.

In Ref. 20 it was demonstrated that electron microscopy has the potential for studying an optical lattice with magnificent resolution, although averaging over many instances of the experiment is currently required. Here, we employ a type of absorption imaging of light to study a 2D optical lattice, an array of lattice sites lying in a horizontal plane. Each lattice site has a slightly elongated cigar shape. For the highest lattice depths reported here, the vertical radius of each site is 2.3 times the horizontal radius. Our lattice can be considered as a hybrid between an array of focused dipole traps[21] and an optical standing wave [1,15,22]. The lattice is a greatly reduced image of an array of foci created by a microlens array, as shown in Fig. 1. However, this image is filtered in the Fourier plane, allowing for a variety of possible lattices. Fig. 1l is an image of the Fourier plane, showing the Fourier transform of the array of foci created by the microlens array. Each of the discrete points in the Fourier plane represents an available lattice beam, which will propagate in a unique direction at the location of the atoms. For our experiment, we use the 4 points circled in blue, giving a lattice period of $a = 2.0$ microns. All of the lattice beams of Fig. 1l have the same frequency. This does not create a stability problem due to slight relative motions of the beams, because the lattice is an image of the microlens array, an object whose shape does not change. In order to verify the uniformity of the lattice beams, we add a 5$^{th}$ beam, circled in green in Fig. 1l. This vertically propagating beam has two effects on the lattice. Firstly, it supports the lattice against gravity by dividing it into planes, and secondly, it increases the lattice period by a factor of $\sqrt{2}$. Fig. 1m,n show a large thermal cloud which has been loaded into the 5-beam lattice, and imaged from two directions. Fig. 1m shows that the lattice is uniform over the region indicated by the green ellipse, which is the region of the 4-beam lattice occupied by the condensate in the rest of this work.

In the 4-beam lattice, an additional laser beam with an elliptical cross section confines the condensate in the horizontal plane, as shown in Fig. 2a,b. We refer to this beam as the vertical confinement beam (Fig. 1k). There is no magnetic trapping potential. The vertical confinement beam has a waist ($1/e^2$ radius) of 3.3 μm in the vertical direction, and 32 μm in the transverse direction. The beam propagates upward in Fig. 2a, and into the image of Fig. 2b. The parabolic minimum of the resulting potential has frequencies of $v_z = 884 \text{ Hz}$ in the vertical direction, $v_x = 91 \text{ Hz}$ in the transverse direction, and $v_y = 34 \text{ Hz}$ in the propagation direction of the beam.

Initially, we create a Bose-Einstein condensate of $1 \times 10^5$ $^{87}$Rb atoms in the $F = 2$, $m_F = 2$ state in a cigar-shaped magnetic trap. The condensate is adiabatically expanded until the trap frequencies are 26 Hz and 10 Hz in the radial and axial directions respectively. The vertical confinement beam is then ramped on adiabatically during 50 msec. The magnetic trapping potential is then ramped off during 200 msec, leaving a purely optical trapping potential (Fig. 2a,b). The lattice is then ramped on, either linearly or exponentially, during a time $\tau_{ramp}$ varying between 10 and 800 msec to a depth $V_o$. For $V_o/h = 0.7$ kHz and chemical potential $\mu/h = 2$ kHz, the lattice only partly divides the condensate into sites, as shown in Fig. 2c,d. For $V_o/h = 7$ kHz and $\mu/h = 3$ kHz, the lattice consists of separate sites where the wavefunction is evanescent between sites, as shown in Fig. 2e,f and Fig. 3a,b. Since the purely optical potential is only 1 µK deep, the atomic sample is extremely cold, and we see no evidence of thermal atoms.

We measure the number fluctuations in each site, by comparing 6 images of the type shown in Fig. 2e,f. We integrate over each site to obtain $N_i$. The average $N_i$ in each of the 6 images varies between 146 and 163. We normalize each image to an average of $N = 154$ atoms per site. We then compute the fluctuations for each site for each image, given by $\delta N_i = N_i - \overline{N}_i$, where $\overline{N}_i$ is averaged over the 6 images. $\delta N_i$ for one of the images is shown in Fig. 4a. No long range order is seen. A histogram of $\delta N_i$ for all 6 images is shown in Fig. 4b. The histogram is seen to be narrower than the Poissonian distribution (dashed line) by a factor of 0.7, so $\sigma_N = 0.7\sqrt{N}$. Both the histogram and the dashed line enclose the same area. The width of the theoretical Poissonian distribution shown in the figure includes three corrections relative to $\sigma_N = \sqrt{154} = 12.4$ atoms, which result from our measurement system. These include a 0.6 atom decrease for the finite resolution of the imaging system, a 0.5 atom increase due to photon shot noise in the imaging, and a 1.1 atom decrease to account for the finite number of images used to compute $\overline{n}_i$. The measured sub-Poissonian distribution indicates that the system exists between the superfluid and Mott insulator regimes, or in the Mott insulator regime. The width of the fluctuation distribution does not decrease below $0.7\sqrt{N}$ due to density-dependent losses.

These losses result from the high densities associated with an optical lattice[4,8,15]. We can understand the high density from Fig. 2e,f where we see that the presence of the lattice potential confines the wavefunction in each site to a significantly smaller volume than in Fig. 2a,b. The losses are clear from the profile of the lattice for various lattice depths, as shown in Fig. 3c,d. The individual lattice sites of Fig. 3c,d have been numerically smoothed away, to better show the overall profile. As the depth of the lattice is increased, the density in each site increases, and the total number of atoms decreases, creating a flat-top profile. This profile indicates that the losses are density-dependent, creating a constant density across the lattice. The flat-top profile results when the loss rate $\tau_m^{-1}$ is larger or comparable to the tunneling rate $J/\hbar$. This criterion is independent of the ramp-up rate $\tau_{ramp}^{-1}$, so even a very slow ramp up does not insure a ground state profile. We verify this by ramping up the lattice in times as long as 800 msec, for which we still observe the flat top profile. Density-dependent losses could possibly explain the losses of ref. 15 as well. Indeed, the average densities in refs. 4, 8, and 15 were 3 to 4 times higher than in our experiment. Fig. 3e,f (also smoothed, except for one trace) shows the decay of a lattice with a smaller initial number of atoms relative to Fig. 3c,d. It is seen that for the smaller atom number, tunneling dominates and the central region of the lattice follows the ground state profile, as opposed to the flat-top profile of Fig. 3c,d. This implies that in general for our system, the tunneling rate $J/\hbar$ is comparable to $\tau_m^{-1}$, so that the change in initial atom number can cause either of these rates to dominate. Thus, the lower limit $\sigma_N \approx \sqrt{5N/[4(m-1)+10]}$ due to density-dependent losses should always play a role in our system.

We have thus found that $J/\hbar$ is on the same order as $\tau_m^{-1}$, so we can measure $\tau_m^{-1}$ in order to obtain an estimate of $J/\hbar$. The inset of Fig 3e shows a semi-log plot of the decay of the central lattice site of Fig. 3e. The decay is non-exponential due to the density dependence of $\tau_m^{-1}$. Assuming that the losses are due to 3-body collisions, $\tau_3^{-1}$ can be deduced from the inset of Fig. 3e. The solid curve shows a fit for 3-body losses, giving $\tau_3^{-1} = (3.2 \times 10^{-29} \text{ cm}^6 \text{ sec}^{-1}) n^2$, where $n^2$ has been averaged over the lattice site. This rate is almost 6 times larger than the rate reported in Ref. 19. Larger than expected decay rates were also reported in Refs. 7, 15, and 18. We take this value of $\tau_3^{-1}$ as an order-of-magnitude estimate of $J/\hbar$, which is consistent with our estimates by Ref. 23.

Using this value of $J/\hbar$, we can find the values of $\sigma_N$ and $\sigma_\phi$ which we would expect in the absence of losses. Extrapolating the predictions of Ref. 24 to large $N$ [17,25], the Mott transition should occur at $U/J = 17N$, which agrees with the predictions of ref. 26. For our lattice sites, $U = d\mu/dN = 2\mu/(5N)$. We thus find $U/J = (8\times 10^{-3})17N$ which gives $\sigma_N = 0.2\sqrt{N}$ and $\sigma_\phi = 0.2$. These widths are relatively insensitive to $J$ since they vary as $J^{\pm 1/4}$. The measured value of $\sigma_N$ is significantly larger due to the losses. Such a small value of $\sigma_\phi$ should produce a clear diffraction pattern, which we observe in the time-of flight images shown in Fig. 3g,i. The reciprocal lattice period $2\pi/a$ is clearly visible in the images. However, such a momentum space image is not sufficient to prove that the phase is uniform[16,27].

These time-of-flight images have a somewhat different appearance than the images in previous works[4]. This is a result of the fact that the momentum spacing between orders, indicated by $2\pi/a$ in Fig. 3g, is smaller than the momentum distribution of the cloud released from the vertical confinement beam in the absence of the lattice, seen in Fig. 3h. Thus, the various orders expand and overlap, forming the pattern of Fig. 3g. Imaging the lattice in time-of-flight from the horizontal direction in Fig. 3i, a reversal is seen in the overall aspect ratio of the momentum distribution when the lattice is turned on, relative to Fig. 3j. This occurs since turning on the lattice reverses the in-situ aspect ratio by converting the horizontal pancake-shaped trap into slightly elongated vertical ellipsoids.

The profile of Fig. 3e,f has implications for the evolution of the phase of the lattice. The profile is seen to be narrower than the ground state profile indicated by blue curves. This ground state profile in the presence of the lattice is significantly wider than the profile with only the harmonic confinement provided by the vertical confining beam. This results from the substantial decrease in the volume per site mentioned above, which is associated with turning on an optical lattice. To avoid a drastic increase in the peak density, the atoms must distribute themselves over a larger number of sites. This should cause the area of the entire cloud in the ground state to increase significantly. This principle is true for almost all lattice experiments. In refs.

4, 8, and 15 for example, in order to remain in the ground state, the turning on of the lattice must have corresponded to an increase in the radius of the cloud by factor of approximately 1.2, 1.7 and 1.7, respectively. Due to the fact that $\tau_{ramp}^{-1}$ is comparable to $J/\hbar$, the lattice of Fig. 3e,f does not follow the entire ground state profile. However, the figure shows that most of the lattice sits on a curve of constant $\mu$. The phase of each lattice site can therefore evolve at almost the same rate, maintaining an extended wavefunction with almost constant phase across the lattice. We thus see that even a lattice which is not in the ground state can have a uniform phase.

In conclusion, we have imaged an optical lattice in-situ, with single-site resolution. We find that losses affect the statistics of the number fluctuations between lattice sites, which can limit the usefulness of the Mott transition to produce a uniform distribution of atoms in a lattice. On the other hand, such losses can be useful for preparing exotic quantum phases[16]. Such losses could possibly be avoided however, by using one or two atoms per site [3,8]. We also find that an almost uniform phase across the lattice does not necessarily imply that the lattice is in the ground state. The in-situ imaging of the current system could be exploited in the study of non-linear effects such as 2D gap solitons[28]. By rotating the microlens array[22], quantum phase transitions of vortex states could also be explored[29-32]. Single-site addressability could readily be added to the system. By combining such a system with single-atom sensitivity, a quantum computer could be created[13].

We thank A. Auerbach, E. Polturak, , R. Pugatch, E. Altman, and Y. Kafri for helpful discussions. This work was supported by the Israel Science Foundation.

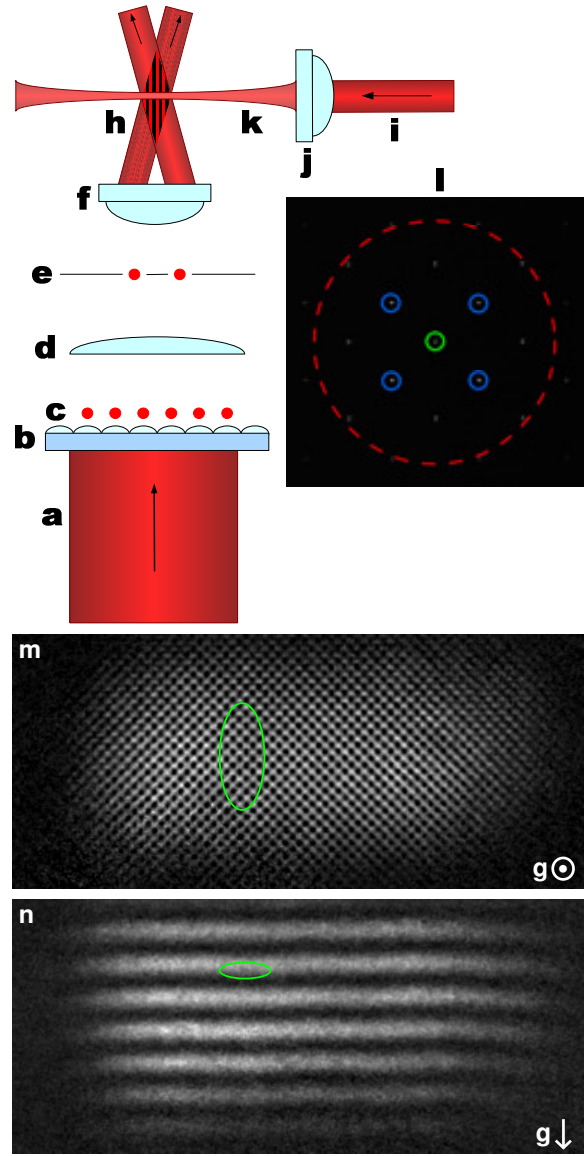

FIG. 1 The experimental system. a, The initial beam. b, The microlens array. c, The array of 2000 foci generated by the microlens array. These foci are similar to delta functions. d, Lens. e, The Fourier plane where the desired lattice beams are selected. f, Aspheric lens used for absorption imaging, as well as producing the lattice. h, The lattice sites, which are an image of c. i, Elliptical beam for the vertical confinement beam. j, Aspheric lens used for absorption imaging, as well as focusing the vertical confinement beam. k, The vertical confinement beam, focused on the lattice. l, An image of the available lattice beams in the Fourier plane e. Each delta function-like point of light will reach the atoms at h from a different direction. The four beams

circled in blue are used in this experiment. The fifth vertically propagating beam circled in green is combined with the other four to provide vertical confinement for diagnostic purposes. The dashed circle indicates the radius of lens f. Only beams within this line are usable. **m,** 5-beam lattice with increased lattice period, imaged through lens f. **n,** the horizontal planes of the 5-beam lattice, as viewed through lens i. Gravity is indicated by "g". The green ellipses in m and n indicate the region of the lattice used for the experiment, which employs 4 lattice beams and the beam k for vertical confinement.

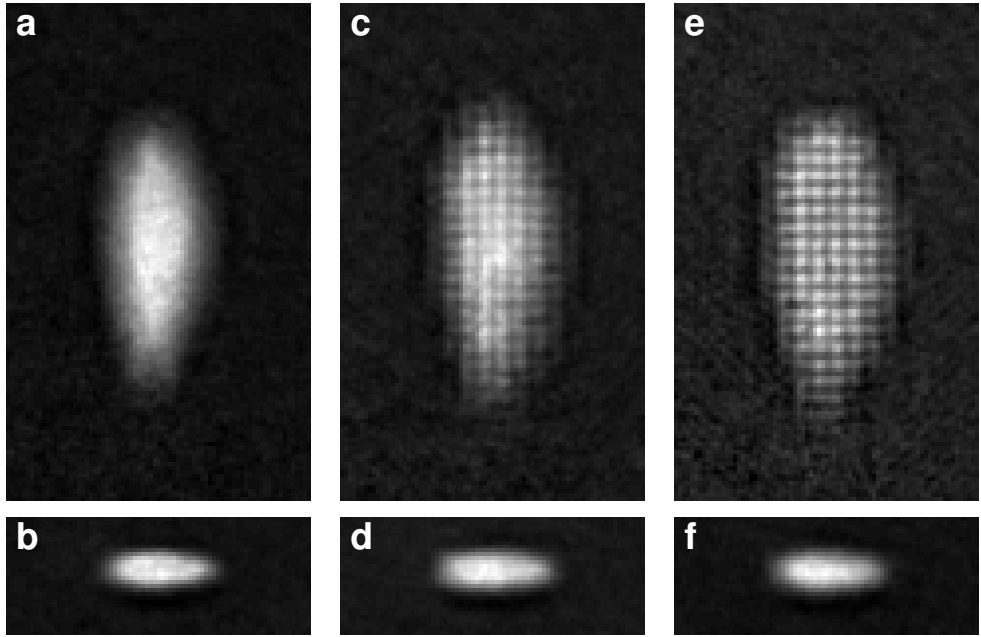

FIG. 2 In-situ images of the lattice, with single-site resolution. a, The condensate in the vertical confinement beam only ($V_o = 0$). The image is in a horizontal plane. b, As in a, but imaged in a vertical plane. c, The lattice with a depth $V_o/h = 0.7$ kHz viewed in a horizontal plane. d, As in c, but imaged in a vertical plane. e, The lattice with a depth $V_o/h = 7$ kHz viewed in a horizontal plane. f, As in e, but imaged in a vertical plane.

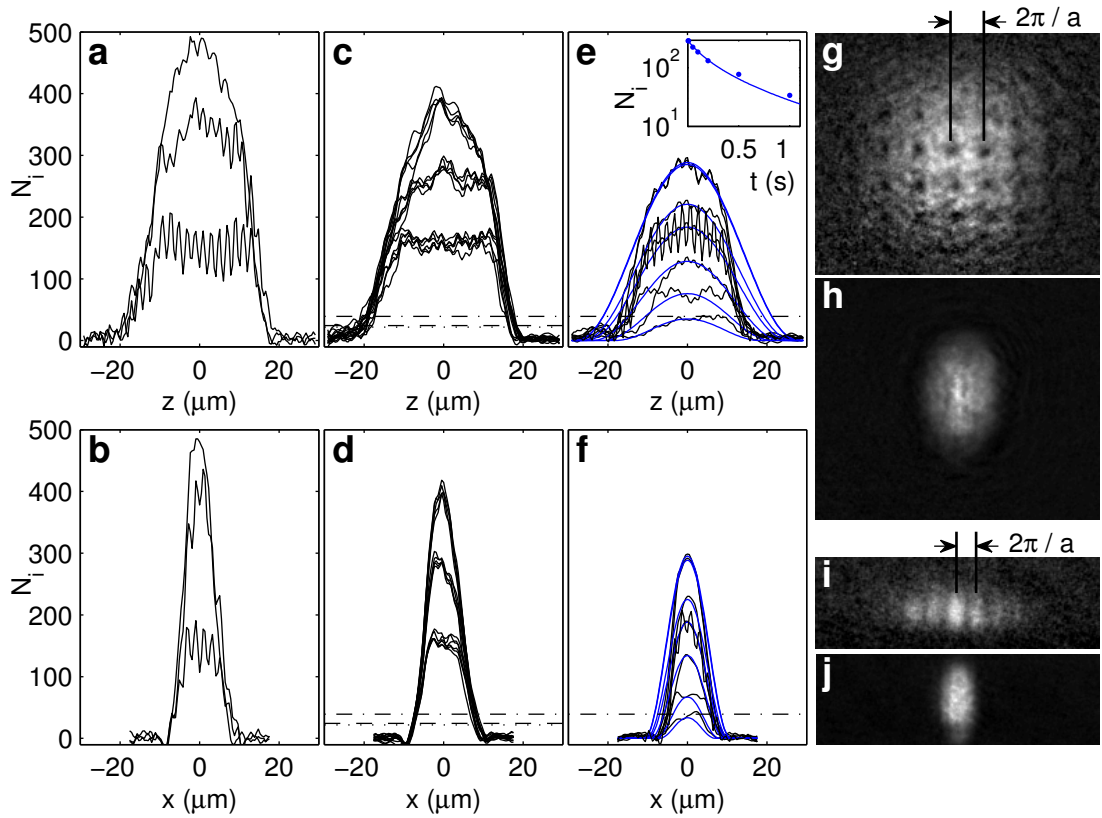

FIG. 3 Profiles of the lattice in-situ, and time-of-flight images. The vertical axes in a-f are in units of atoms per lattice site. **a,** profiles of the lattice in the *y*-direction, summed over the pixels composing the central column of lattice sites. The upper, middle, and lower traces correspond to Figs. 2a, 2c, and 2e, respectively. **b,** profiles in the *x*-direction, corresponding to the profiles in the *y*-direction shown in a. **c,** Smoothed *y* profiles of the lattice. The individual lattice sites have been smoothed away to better show the overall profile. The upper, middle, and lower profiles correspond to $V_o/h$ = 0.7 kHz, 1.0 kHz, and 7 kHz, respectively. Several profiles for each depth are shown. Below the horizontal lines, the wavefunction of each lattice site can be considered to be one-dimensional (1D)[33,34]. Above the lines, the wavefunction is 3D. The dash-dotted line corresponds to the lower profiles, the dashed line corresponds to the middle profiles, and the dotted line corresponds to the upper profiles. **d,** *x* profiles corresponding to the *y* profiles of c. **e,** Smoothed *y* profiles showing the decay of the lattice after a 10 ms ramp-up to $V_o/h$ = 7 kHz. One profile is shown with and without smoothing. Times between 5 and 1000 ms after the ramp-up are shown. The blue curves show the theoretical ground state

profiles. The dash-dotted line shows the transition from 1D to 3D lattice sites. The inset shows the population of the central lattice site from each of the profiles. The solid curve of the inset shows a fit taking one and 3-body losses into account. The 3-body loss rate is an adjustable parameter in the fit. **f,** $x$ profiles corresponding to the $y$ profiles of e. **g,** The lattice after 5 msec time-of-flight, imaged in a horizontal plane. **h,** 5 msec time-of-flight from the vertical confinement beam only ($V_o = 0$). **i, j,** Same as g,h but with 3 msec time-of-flight, viewed in a vertical plane.

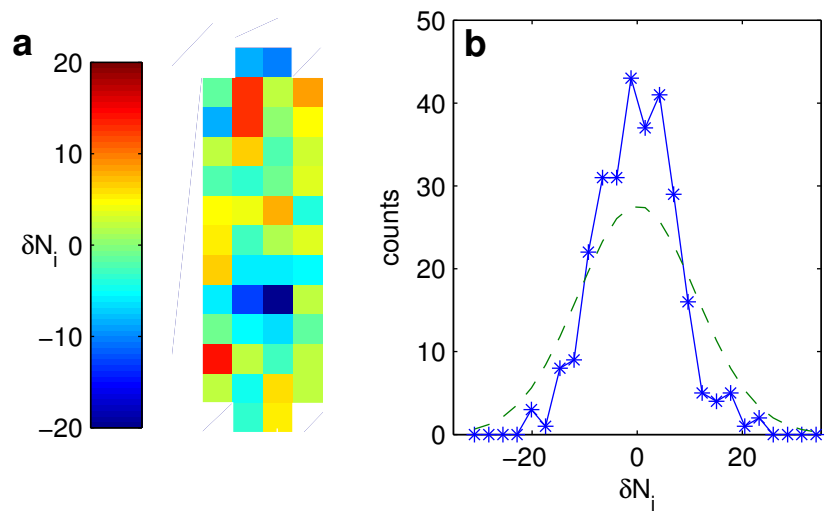

FIG. 4 In-situ observation of sub-Poissonian number fluctuations in individual lattice sites. a, One instance of the optical lattice, showing the population in each site, relative to the mean population for each site. The color scale indicates the number of atoms. b, The stars and the solid curve show a histogram of the fluctuations for 6 images like a. The dashed curve indicates the Poissonian distribution. The vertical axis indicates the total number of sites in 6 images with a given value of $\delta N_i$.

1. Anderson, B. P. & Kasevich, M. A. Macroscopic quantum interference from atomic tunnel arrays. *Science* **282**, 1686 (1998).

2. Fisher, M. P. A.,Weichman, P. B., Grinstein, G. & Fisher, D. S. Boson localization and the superfluid-insulator transition. *Phys. Rev. B* **40**, 546-570 (1989).

3. Jaksch, D., Bruder, C., Cirac, J. I., Gardiner, C. W. & Zoller, P. Cold bosonic atoms in optical lattices. *Phys. Rev. Lett.* **81**, 3108 (1998).

4. Greiner, M., Mandel, O., Esslinger, T., Hänsch, T. W. & Bloch, I. Quantum phase transition from a superfluid to a Mott insulator in a gas of ultracold atoms. *Nature* **415**, 39 (2002).

5. Spielman, I. B., Phillips, W. D. & Porto, J.V. Mott-insulator transition in a two-dimensional atomic Bose gas. *Phys. Rev. Lett.* **98**, 080404 (2007).

6. Orzel, C., Tuchman, A. K., Fenselau, M. L., Yasuda, M. & Kasevich, M. A. Squeezed States in a Bose-Einstein Condensate. *Science* **291**, 2386 (2001).

7. Gerbier. F., Fölling, S., Widera, A., Mandel, O. & Bloch, I. Probing number squeezing of ultracold atoms across the superfluid-Mott insulator transition. *Phys. Rev. Lett.* **96**, 090401 (2006).

8. Campbell, G. K., Mun, J., Boyd, M., Medley, P., Leanhardt, A. E., Marcassa, L. G., Pritchard, D. E. & Ketterle, W. Imaging the Mott insulator shells by using atomic clock shifts. *Science* **313**, 649 (2006).

9. Fölling, S., Widera, A., Müller, T., Gerbier. F. & Bloch, I. Formation of Spatial Shell Structure in the Superfluid to Mott Insulator Transition. *Phys. Rev. Lett.* **97**, 060403 (2006).

10. Mun, J., Medley, P., Campbell, G. K., Marcassa, L. G., Pritchard, D. E. & Ketterle, W. Phase diagram for a Bose-Einstein condensate moving in an optical lattice. *Phys. Rev. Lett.* **99**, 150604 (2007).

11. Morsch, O. & Oberthaler, M. Dynamics of Bose-Einstein condensates in optical lattices. *Rev. Mod. Phys.* **78**, 179-215 (2006).

12. Bloch, I., Dalibard, J. & Zwerger, W. Many-body physics with ultracold gases. *Rev. Mod. Phys.* **80**, 885-964 (2008).

13. Jaksch, D. Briegel, H.-J., Cirac, J. I., Gardiner, C. W. & Zoller, P. Entanglement of atoms via cold controlled collisions. *Phys. Rev. Lett.* **82**, 1975-1978 (1999).

14. Schneider, U., Hackermüller, L., Will, S., Best, Th., Bloch, I., Costi, T. A., Helmes, R. W., Rasch, D. & Rosch, A. Metallic and insulating phases of repulsively interacting fermions in a 3D optical lattice. Science **322**, 1520-1525 (2008).

15. Greiner, M., Bloch, I., Mandel, O., Hänsch, T. W. & Esslinger, T. Exploring phase coherence in a 2D lattice of Bose-Einstein condensates. *Phys. Rev. Lett.* **87**, 160405 (2001).

16. Daley, A. J., Taylor, J. M., Diehl, S., Baranov, M. & Zoller, P., Atomic three-body loss as a dynamical three-body interaction. *Phys. Rev. Lett.* **102**, 040402 (2009).

17. Javanainen, J. Phonon approach to an array of traps containing Bose-Einstein condensates. *Phys. Rev. A* **60**, 4902-4909 (1999).

18. Hadzibabic, Z., Stock, S., Battelier, B., Bretin, V. & Dalibard, J., Interference of an array of independent Bose-Einstein condensates. *Phys. Rev. Lett.* **93**, 180403 (2004).

19. Burt, E. A., Ghrist, R. W., Myatt, C. J., Holland, M. J., Cornell, E. A. & Wieman, C. E. Coherence, correlations, and collisions: What one learns about Bose-Einstein condensates from their decay. *Phys. Rev. Lett.* **79**, 337-340 (1997).

20. Gericke, T., Würtz, P., Reitz, D., Langen, T. & Ott, H. High-resolution scanning electron microscopy of an ultracold quantum gas. *Nature Phys.* **4**, 949-953 (2008).

21. Dumke, R., Volk, M., Müther, T., Buchkremer, F. B. J., Birkl, G. & Ertmer W. Micro-optical realization of arrays of selectively addressable dipole traps: a scalable configuration for quantum computation with atomic qubits. *Phys. Rev. Lett.* **89**, 097903 (2002).

22. Tung, S., Schweikhard, V. & Cornell, E. A. Observation of vortex pinning in Bose-Einstein condensates. *Phys. Rev. Lett.* **97**, 240402 (2006).

23. Dalfovo, F., Pitaevskii, L. & Stringari, S. Order parameter at the boundary of a trapped Bose gas. *Phys. Rev. A* **54**, 4213-4217 (1996).

24. Elstner, N. & Monien, H. Dynamics and thermodynamics of the Bose-Hubbard model. *Phys. Rev. B* **59**, 12184-12187 (1999).

25. Javanainen, J. & Ivanov, M. Y. Splitting a trap containing a Bose-Einstein condensate: Atom number fluctuations. *Phys. Rev. A* **60**, 2351-2359 (1999).

26. van Oosten, D., van der Straten, P. & Stoof, H. T. C. Mott insulators in an optical lattice with high filling factors. *Phys. Rev. A* **67**, 033606 (2003).

27. Fallani, L., Fort, C., Lye, J. E. & Inguscio M. Bose-Einstein condensate in an optical lattice with tunable spacing: transport and static properties. *Optics Express* **13**, 4303-4313 (2005).

28. Ostrovskaya, E. A. & Kivshar, Y. S. Matter-wave gap solitons in atomic band-gap structures. *Phys. Rev. Lett.* **90**, 160407 (2003).

29. Pu, H., Baksmaty, L. O., Yi, S. & Bigelow, N. P. Structural phase transitions of vortex matter in an optical lattice. *Phys. Rev. Lett.* **94**, 190401 (2005).

30. Wu, C., Chen, H.-D., Hu, J.-P. & Zhang, S.-C. Vortex configurations of bosons in an optical lattice. *Phys. Rev. A* **69**, 043609 (2004).

31. Bhat, R., Holland, M. J. & Carr, L. D. Bose-Einstein condensates in rotating lattices. *Phys. Rev. Lett.* **96**, 060405 (2006).

32. Reijnders, J. W. & Duine, R. A. Pinning and collective modes of a vortex lattice in a Bose-Einstein condensate. *Phys. Rev. A* **71**, 063607 (2005).

33. Görlitz, A., Vogels, J. M., Leanhardt, A. E., Raman, C., Gustavson, T. L., Abo-Shaeer, J. R., Chikkatur, A. P., Gupta, S., Inouye, S., Rosenband, T. & Ketterle, W. Realization of Bose-Einstein condensates in lower dimensions. *Phys. Rev. Lett.* **87**, 130402 (2001).

34. Pitaevskii, L. & Stringari, S. *Bose-Einstein Condensation* Sect. 17.2 (Oxford Univ. Press, 2003).

# Appendix

### The uncertainty in the population due to stochastic losses

Consider a lattice with exactly $N$ atoms per site at time $t = 0$. The sites are completely isolated from one another. The sites are subject to stochastic $m$-body losses with a rate $\tau_m^{-1}$. Due to the stochastic nature of the losses, a population difference between sites will develop. However, this growth in the fluctuations will be balanced by the decay in $N$. To see this, consider two sites $A$ and $B$ within the lattice. If at a given time, site $A$ has a larger population than site $B$, then site $A$ will lose atoms at a higher rate than site $B$. Furthermore, $\tau_m^{-1}$ might be larger for a larger population, which will further limit the growth of the fluctuations.

First, let us consider the case of constant $\tau_m^{-1}$. After a time $t$ which is small compared with $\tau_m$, the average number of atoms lost per site will be $Nt/\tau_m$. The fluctuation in the population of a given site at time $t$ will thus reach

$$\delta N = \sqrt{Nt/\tau_m}. \qquad (1)$$

Therefore, $\delta N$ will continually grow at a rate given by

$$\frac{d\delta N}{dt} = \frac{N}{2\tau_m \delta N}. \qquad (2)$$

However, the growth of $\delta N$ changes the rate of growth $N\tau_m^{-1}(N)$. Specifically, if at time $t$ the population is $N + \delta N$ (a larger population than expected), then the increase in the loss rate will be $\Delta(N\tau_m^{-1}) = (N + \delta N)\tau_m^{-1}(N + \delta N) - N\tau_m^{-1}(N)$. To first order in $\delta N / N$, this can be written as

$$\Delta(N\tau_m^{-1}) = N \frac{d\tau_m^{-1}}{dN}\delta N + \delta N \tau_m^{-1} \qquad (3)$$

At the equilibrium value of $\delta N$, the rates (2) and (3) will be equal. Note that equating (2) and (3) is not self-consistent, since the former was derived with constant $N\tau_m^{-1}$, and the latter with variable $N\tau_m^{-1}$. Equating them nevertheless gives the approximate result:

$$\delta N^2 \approx \frac{N}{2}\left(N\tau_m \frac{d\tau_m^{-1}}{dN} + 1\right)^{-1}. \qquad (S4)$$

To apply (4) to our system, we should write down the dependence of $\tau_m^{-1}$ on $N$. Specifically, $\tau_m^{-1} \propto \langle n^{m-1} \rangle$, where $n$ is the density in the site, and the average is taken over all of the atoms in the site. For our lattice sites, $\langle n^{m-1} \rangle \propto N^{2(m-1)/5}$. Thus,

$$\tau_m^{-1} = CN^{2(m-1)/5} \qquad (S5)$$

where $C$ is a constant. Plugging (5) into (4) yields

$$\delta N \approx \sqrt{\frac{5N}{4(m-1)+10}} \ . \tag{S6}$$

For 3-body losses, (6) gives $\delta N \approx \sqrt{5N/18} = 0.53\sqrt{N}$. A numerical simulation of the stochastic decay process gives a value which is larger by roughly 15%.